\documentclass[aps,prl,preprint,groupedaddress]{revtex4}

\usepackage{graphics}

\begin{document}


\title{Optimization of qPlus sensor geometry and circuit for high-speed atomic force microscopy in liquid environments}


\author{Takashi Ichii}
\email{ichii.takashi.2m@kyoto-u.ac.jp}
\author{Shuji Tokitoh}
\author{Yuto Nishiwaki}
\author{Toru Utsunomiya}

\affiliation{Department of Materials Science and Engineering, Kyoto University, Yoshida-hommachi, Sakyo, Kyoto 606-8501, Japan}


\date{\today}

\begin{abstract}

Atomic force microscopy (AFM) using qPlus sensors is a powerful tool for high-resolution analysis in various liquids, including high-viscosity or opaque environments. 
However, the relatively high displacement sensor noise density ($n_{\mathrm{ds}}$), combined with the high spring constant and the low resonance frequency, limits force sensitivity and has hindered high-speed imaging. 
In this paper, we clarify the dominant factors governing $n_{\mathrm{ds}}$ and the minimum detectable force gradient ($F'_{\mathrm{min}}$) through a comprehensive analysis of sensor geometry and circuit theory. 
Based on these findings, we developed a low-noise qPlus sensor that achieves an $n_{\mathrm{ds}}$ of $9.3 \mathrm{~fm~Hz^{-1/2}}$, which is approximately one-third that of conventional sensors, and reduces $F'_{\mathrm{min}}$ by half. 
Using this sensor, we demonstrated high-speed, atomic-resolution imaging of a molten gallium interface at a frame rate of $6.6 \mathrm{~s~frame^{-1}}$ (39 lines $\mathrm{s^{-1}}$), proving its advantage for analyzing fast interfacial dynamics in liquid environments.

\end{abstract}


\maketitle




Atomic force microscopy (AFM) is a powerful technique for analyzing the structure of solid-liquid interfaces at the atomic and molecular scales. 
In many liquid-environment AFM systems, silicon microcantilevers are employed as force sensors, with their deflection detected by optical methods such as the optical beam deflection or interferometry\cite{Fukuma05,Fukuma05_2,Hoogenboom06,Rasool10}. 
However, these methods are difficult to apply to optically opaque or highly viscous liquids. 
To address these issues, we have utilized the qPlus sensor as a force sensor to achieve atomic and molecular scale AFM analysis in high-viscosity liquids, such as ionic liquids and polymer melts, as well as in opaque molten metals\cite{Ichii12,Ichii14,Yamada20,Nishiwaki24,Ichii21}. 
Thus, liquid-environment AFM using the qPlus sensor is highly effective for high-resolution analysis in environments where conventional silicon microcantilevers are inapplicable.

On the other hand, the qPlus sensor typically exhibits a lower resonance frequency ($f$) and a higher spring constant ($k$) compared to silicon microcantilevers\cite{Giessibl98,Giessibl00}. 
These characteristics lead to two primary issues. First, the low resonance frequency limits the measurement speed. 
The qPlus sensor is generally operated in frequency modulation AFM (FM-AFM) using a phase-locked loop (PLL) for sensor excitation and frequency shift detection\cite{Durig97,Kobayashi01}. 
Since the PLL bandwidth is generally restricted to approximately $1/20$ of the resonance frequency, and the resonance frequency of a typical qPlus sensor is between $15$ and $30\mathrm{~kHz}$, its operational bandwidth has historically been limited to roughly $1\mathrm{~kHz}$. 
Second, the relatively high displacement sensor noise density ($n_{\mathrm{ds}}$), combined with the low resonance frequency and the high spring constant, results in an increase in the minimum detectable force gradient ($F'_{\mathrm{min}}$)\cite{Albrecht91}. 
The reduction in measurement speed caused by these two issues hinders the observation of various dynamics at the solid-liquid interface.

Recently, we developed a hybrid-loop frequency demodulator that successfully extended the bandwidth to over $5\mathrm{~kHz}$, addressing the first of the aforementioned issues\cite{Nishiwaki26}. 
However, the second issue regarding force sensitivity remains unresolved. 
In this study, we theoretically clarify the factors determining $F'_{\mathrm{min}}$ in qPlus sensors and, based on these findings, develop a low-noise qPlus sensor. 
Using this sensor, we demonstrate high-speed, atomic-resolution AFM analysis in a liquid environment.

\begin{figure*}[htb]
\includegraphics{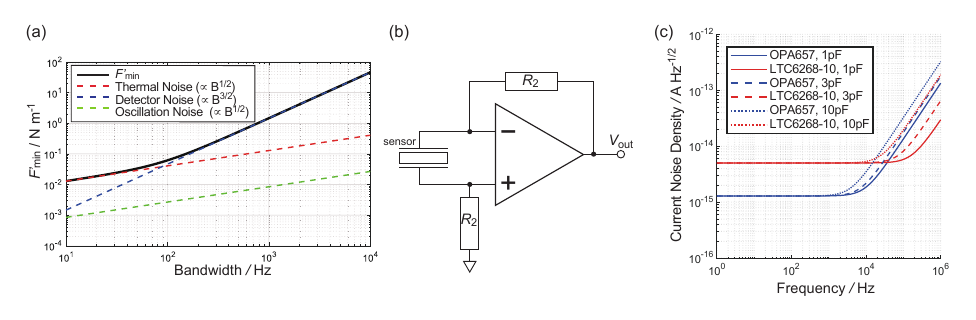}
\caption{\label{Fig_1} (a) Calculated minimum detectable force gradient ($F'_{\mathrm{min}}$) of a qPlus sensor under typical conditions: $k = 1890 \mathrm{~N~m^{-1}}$, $T = 300 \mathrm{~K}$, $A = 100 \mathrm{~pm}$, $f = 30 \mathrm{~kHz}$, $n_{\mathrm{ds}} = 30 \mathrm{~fm~Hz^{-1/2}}$, and $Q = 3000$. (b) Differential transimpedance amplifier (TIA) proposed by Huber $et$ $al.$\cite{Huber17}. (c) Frequency dependence of current noise density ($I_{\mathrm{n}}$) of the differential TIA with various input capacitance.}
\end{figure*}

$F'_{\mathrm{min}}$ in FM-AFM under the assumption of small amplitude is given by the following equation\cite{Albrecht91,Kobayashi09}:
\begin{eqnarray}
F'_{\mathrm{min}}
= \sqrt{\frac{4kk_{B}TB}{\pi A^{2}fQ} + \frac{8k^{2}n_{\mathrm{ds}}^{2}B^{3}}{3 A^{2} f^{2}}+\frac{2k^2n_{\mathrm{ds}}^2B}{A^2Q^2}} 
\end{eqnarray}
Here, $k$ is the spring constant ($\mathrm{N~m^{-1}}$), $k_{\mathrm{B}}$ is the Boltzmann constant, $T$ is the temperature ($\mathrm{K}$), $B$ is the bandwidth ($\mathrm{Hz}$), $A$ is the oscillation amplitude ($\mathrm{m}$), $f$ is the resonance frequency of the sensor ($\mathrm{Hz}$), $n_{\mathrm{ds}}$ is the displacement sensor noise density ($\mathrm{fm~Hz^{-1/2}}$), and $Q$ is the quality factor. In Eq. (1), the first term under the square root represents the thermal noise, the second term represents the displacement sensor noise, and the third term represents the oscillator noise.
Figure 1(a) shows the calculated $F'_{\mathrm{min}}$ under typical conditions: $k = 1890 \mathrm{~N~m^{-1}}$, $T = 300 \mathrm{~K}$, $A = 100 \mathrm{~pm}$, $f = 30 \mathrm{~kHz}$, $n_{\mathrm{ds}} = 30 \mathrm{~fm~Hz^{-1/2}}$, and $Q = 3000$. The results indicate that while thermal noise is dominant for $B < \sim100$ Hz, displacement sensor noise becomes dominant for $B > \sim1$ kHz. The oscillator noise has almost no contribution in any frequency range. 
Therefore, reducing the displacement sensor noise is crucial for high-speed AFM measurements.

Since the displacement sensor noise is proportional to $n_{\mathrm{ds}}$, it is necessary to reduce $n_{\mathrm{ds}}$ to minimize the displacement sensor noise. 
Here, $n_{\mathrm{ds}}$ is defined by the voltage noise density $E_{\mathrm{n}}$  ($\mathrm{V~Hz^{-1/2}}$) and the detection sensitivity $S_{\mathrm{v}}~(\mathrm{V~m^{-1}})$ as $n_{\mathrm{ds}} = E_{\mathrm{n}} / S_{\mathrm{v}}$. 
The quartz tuning fork (QTF) that constitutes the qPlus sensor is a piezoelectric element, and its vibration is detected by a transimpedance amplifier (TIA). 
In particular, the differential TIA proposed by Huber $et$ $al$. (Figure 1(b)) is widely used\cite{Huber17}. 
The resistance of of the two resistors in this TIA ($R_2$) is typically $10~\mathrm{G\Omega}$. 
Assuming a typical parasitic capacitance of these resistors of $0.1~\mathrm{pF}$, the cutoff frequency of the TIA is approximately $160~\mathrm{Hz}$. 
Since this is sufficiently lower than the resonance frequency of the qPlus sensor, the TIA gain $G$ can be expressed as $G = G_0 / f  ~(\mathrm{V~A^{-1}}) $. 
Here, the relationship between the periodic charge fluctuation $Q = Q_0 \exp(2 \pi i f t)~(\mathrm{C})$ and the current $I~(\mathrm{A})$ is given by:
\begin{eqnarray}
I = \dot{Q} = 2 \pi i f Q_0 \exp(2 \pi i f t)
\end{eqnarray}
That is,
\begin{eqnarray}
S_{\mathrm{v}} = \frac{V_0}{A} = \frac{2 \pi f Q_0 G}{A} = \frac{2 \pi Q_0 G_0}{A} = 2 \pi G_0 S_{\mathrm{q}}
\end{eqnarray}
Given that $E_{\mathrm{n}} = G I_{\mathrm{n}} = G_0 I_{\mathrm{n}} / f$, we obtain:
\begin{eqnarray}
n_{\mathrm{ds}} = \frac{I_{\mathrm{n}}}{2 \pi f S_{\mathrm{q}}} \label{eq:nds}
\end{eqnarray}
Here, $S_{\mathrm{q}}$ $(\mathrm{C~m^{-1}})$ is the piezoelectric sensitivity, and $I_{\mathrm{n}}$ ($\mathrm{A~Hz^{-1/2}}$) is the current noise density. 
In other words, to reduce $n_{\mathrm{ds}}$, one must decrease $I_{\mathrm{n}}$ and increase both $f$ and $S_{\mathrm{q}}$. 
The current noise $I_{\mathrm{n}}$ of a conventional (single-ended) TIA is given by the following equation\cite{TI_SBOA570,Eremeev24}:
\begin{eqnarray}
I_{\mathrm{n}} = \sqrt{ \left| \left( \frac{1}{Z_{1}} + \frac{1}{Z_{2}} \right) e_{\mathrm{n}} \right|^2 + |i_{\mathrm{n}}|^2 + \frac{4 k_{\mathrm{B}} T}{R_{2}} }
\end{eqnarray}
where $Z_{1}$ is the input impedance, $Z_{2}$ and $R_{2}$ are the impedance and the resistance of the feedback resistor, respectively, and $e_{\mathrm{n}}$ and $i_{\mathrm{n}}$ are the voltage noise density and current noise density of the operational amplifier, respectively.
Furthermore, a differential TIA can reduce the noise to $1/\sqrt{2}$ compared to a conventional TIA\cite{Huber17}; therefore, $I_{\mathrm{n}}$ can be expressed as follows:
\begin{eqnarray}
I_{\mathrm{n}} = \frac{1}{\sqrt{2}} \sqrt{ \left| \left( \frac{1}{Z_{1}} + \frac{1}{Z_{2}} \right) e_{\mathrm{n}} \right|^2 + |i_{\mathrm{n}}|^2 + \frac{4 k_{\mathrm{B}} T}{R_{2}} }
\end{eqnarray}
Figure 1(c) shows the frequency dependence of $I_{\mathrm{n}}$ using OPA657 (Texas Instruments Inc.) and LTC6268-10 (Analog Devices Inc.) as examples, both of which are known as high-speed, low-noise operational amplifiers for TIAs. Based on their respective datasheets, the values for $e_{\mathrm{n}}$, $i_{\mathrm{n}}$, input resistance, and input capacitance were set as follows: $4.8 \mathrm{~nV~Hz^{-1/2}}$, $1.3 \mathrm{~fA~Hz^{-1/2}}$, $10^{12} \mathrm{~\Omega}$, and $5.2 \mathrm{~pF}$ for OPA657\cite{OPA657}, and $4.0 \mathrm{~nV~Hz^{-1/2}}$, $7.0 \mathrm{~fA~Hz^{-1/2}}$, $10^{12} \mathrm{~\Omega}$, and $0.55 \mathrm{~pF}$ for LTC6268-10\cite{LTC6268-10}. 
It should be noted that $e_{\mathrm{n}}$ and $i_{\mathrm{n}}$ generally depend on frequency; if these values for the operating frequency range of the qPlus sensor were not explicitly provided in the datasheets, the values at the nearest available frequency were adopted. Furthermore, we compared cases where the extrinsic input capacitance $C_{1}$, which is originating from both the sensor and the circuit layout, was $1 \mathrm{~pF}$, $3 \mathrm{pF}$, and $10 \mathrm{pF}$. 
The total capacitance at the input node is the sum of these $C_{1}$ values and the intrinsic input capacitance of the operational amplifier itself.

First, for both OPA657 and LTC6268-10, $I_{\mathrm{n}}$ remains constant and independent of $C_{1}$ in the low-frequency region ($f < \sim1 \mathrm{~kHz}$), indicating that $i_{\mathrm{n}}$ is the dominant noise source in this range. 
The fact that OPA657 exhibits a lower $I_{\mathrm{n}}$ than LTC6268-10 at low frequencies is attributed to the difference in their $i_{\mathrm{n}}$ values.
Furthermore, for both amplifiers, a larger $C_{1}$ causes $I_{\mathrm{n}}$ to begin increasing at lower frequencies, and a larger $C_{1}$ results in a higher $I_{\mathrm{n}}$ overall. 
This implies that the crossover frequency from the $i_{\mathrm{n}}$-dominant to the $e_{\mathrm{n}}$-dominant regime depends on $C_{1}$, and that reducing $C_{1}$ effectively suppresses the influence of $e_{\mathrm{n}}$. 
Notably, above approximately $10 \mathrm{~kHz}$, LTC6268-10 shows a lower $I_{\mathrm{n}}$ than OPA657 for all $C_{1}$ values, which stems from the fact that the input capacitance of LTC6268-10 is only about one-tenth that of OPA657.
The resonance frequency of qPlus sensors, which typically ranges from $15$ to $30 \mathrm{~kHz}$ depending on the tip mass, falls exactly within the transition region from $i_{\mathrm{n}}$-dominant to $e_{\mathrm{n}}$-dominant and where LTC6268-10 provides superior noise performance. 
Therefore, these results clearly demonstrate that employing LTC6268-10 while strictly minimizing $C_{1}$ is highly effective for reducing the overall noise.
Note that, the Johnson noise from the differential TIA configuration (approximately $\sqrt{4k_BT/R_2}/\sqrt{2}$) is about $1 \mathrm{~fA~Hz^{-1/2}}$ under all conditions, which is almost negligible on $I_{\mathrm{n}}$.

To reduce $C_{1}$, we adopted the following approach in this study. 
Conventional qPlus sensors are typically fabricated by fixing one prong of a QTF to an alumina substrate featuring through-hole electrodes using an adhesive. 
However, the high relative permittivity of alumina ($\epsilon \simeq 10$), combined with the through-hole geometry, leads to parasitic capacitive coupling between the electrodes. 
To suppress this, we employed a porous glass substrate with a significantly lower dielectric constant ($\epsilon <$ 2.5 \cite{Cao25}, purchased from Akagawa Glass Co., Ltd., with a porosity of approximately $49\%$), and established direct wiring on the substrate surface instead of utilizing through-holes. 
Furthermore, the epoxy adhesive used to fix the QTF was changed from the conventionally used H74 (purchased from Epoxy Technology, $\epsilon \simeq 4.95$\cite{H74}) to 323LP (also from Epoxy Technology, $\epsilon \simeq 2.62$\cite{323LP}).

\begin{figure*}[htb]
\includegraphics{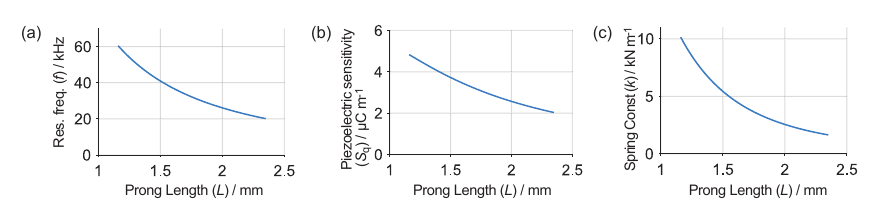}
\caption{\label{Fig_2} Calculated prong length ($L$) dependence of (a) resonance frequency ($f$), (b) piezoelectric sensitivity ($S_{\mathrm{q}}$), and (c) spring constant ($k$) of the qPlus sensor. The prong width $b$, thickness $h$, and electrode length $L_{\mathrm{e}}$ are $0.111 \mathrm{~mm}$, $0.223 \mathrm{~mm}$, and $1.16 \mathrm{~mm}$, respectively. The diameter and length of the tungsten tip are $0.10 \mathrm{~mm}$ and $0.60 \mathrm{~mm}$, respectively.}
\end{figure*}

Next, the increase of the resonance frequency $f$ and the piezoelectric sensitivity $S_{\mathrm{q}}$ are discussed. 
There are three possible approaches to increasing $f$ of a qPlus sensor: (1) utilizing higher-order resonance modes, (2) reducing the mass load on the QTF by shortening the tip, and (3) mechanically shortening the QTF prongs to increase the resonance frequency. 
Approach (1) is unsuitable for vertical force detection because the oscillation direction of the tip shifts toward the horizontal direction in higher-order modes\cite{Yamada23}. 
Approach (2) is practically difficult to implement in liquid-environment AFM applications, where the tip apex is immersed in a droplet. 
In addition, this approach can only increase the frequency up to 32.768 kHz, which is the original resonance frequency of the QTF itself.
Therefore, in this study, we adopted approach (3), the shortening of the QTF prongs. 
Although the QTF is coated with electrodes to extract the piezoelectric current, there are no electrodes near the ends of the prongs; thus, cutting this portion does not affect the detection of the piezoelectric current. 
Furthermore, $S_{\mathrm{q}}$ is given by the following equation\cite{Tung10,Yamada23}:
\begin{eqnarray}
S_{\mathrm{q}} &=&-\frac{hE_{q}d_{31}}{A} \int_{0}^{L_{\mathrm{e}}} \int_{0}^{b} \frac{\partial ^2 \Phi \left( x \right) }{\partial x^2}dydx\\
&=&-\frac{bhE_{q}d_{31}}{A} \left.\frac{\partial \Phi \left( x\right) }{\partial x}\right|_{x = L_{e}}
\end{eqnarray}
Here, $E_{\mathrm{q}}$ is the Young's modulus of quartz ($80 \mathrm{~GPa}$), $d_{31}$ is the piezoelectric constant of quartz ($2.31 \mathrm{~pC~N ^{-1}}$), $b$ and $h$ are the width and thickness of the QTF prong, respectively, $x$ is the longitudinal coordinate along the QTF prong measured from the fixed end ($x = 0$) to the free end ($x = L$), and $L_{\mathrm{e}}$ is the electrode length. 
$\Phi(x)$ is the eigenmode function (or mode shape) of the QTF prong, representing the spatial distribution of the displacement $w(x, t) = \Phi(x) \exp(2\pi i ft)$, and $\Phi(L)$ corresponds to the oscillation amplitude $A$.
That is, $S_{\mathrm{q}}$ is proportional to the slope at the end of the electrode. 
Therefore, for the same oscillation amplitude $A$, a shorter prong results in a larger slope, thereby increasing $S_{\mathrm{q}}$.
However, if the prong is shortened while keeping $b$ and $h$ constant, the spring constant $k$ also increases. 
As shown in Eq. (1), this leads to an increase in $F'_{\mathrm{min}}$. 
Therefore, we calculated $f$, $S_{\mathrm{q}}$, and $k$ as a function of the prong length according to Ref. \cite{Yamada23}. 
The results are shown in Figs. 2(a)-(c).
In this study, we used a QTF purchased from SII Crystal Technology Inc. 
The prong length $L$, width $b$, thickness $h$, and electrode length $L_{\mathrm{e}}$, measured using an optical microscope, were $2.353 \mathrm{~mm}$, $0.111 \mathrm{~mm}$, $0.223 \mathrm{~mm}$, and $1.16 \mathrm{~mm}$, respectively. 
Based on these dimensions, we calculated $f$, $S_{\mathrm{q}}$, and $k$ while varying $L$ from $1.16 \mathrm{~mm}$ to $2.353 \mathrm{~mm}$. 
The diameter and length of the tungsten tip used in the calculations were $0.10 \mathrm{~mm}$ and $0.60 \mathrm{~mm}$, respectively.
The results show that $f$, $S_{\mathrm{q}}$, and $k$ all increase as $L$ decreases. 

\begin{figure*}[htb]
\includegraphics{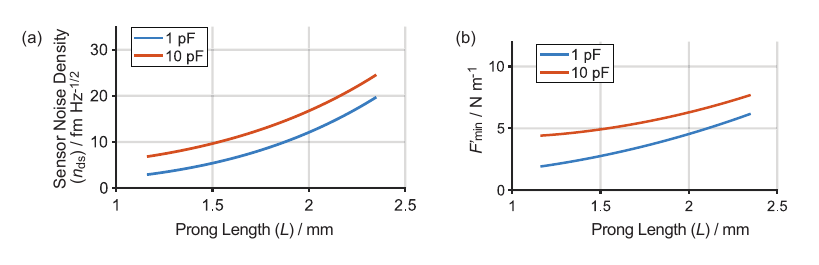}
\caption{\label{Fig_3} Calculated prong length ($L$) dependence of (a) displacement sensor noise density ($n_{\mathrm{ds}}$) and (b)  minimum detectable force gradient ($F'_{\mathrm{min}}$) of the qPlus sensor with a extrinsic input capacitance ($C_1$) of 1 pF and 10 pF. The other parameters of the qPlus sensor are the same as those in Figure 2.}
\end{figure*}

To clarify how these parameters contribute to $n_{\mathrm{ds}}$ and $F'_{\mathrm{min}}$, we calculated the $L$-dependence of $n_{\mathrm{ds}}$ and $F'_{\mathrm{min}}$ for two different conditions, $C_{1} = 1 \mathrm{~pF}$ and $10 \mathrm{~pF}$, according to Eqs. (4) and (1), respectively. 
For the calculation of $F'_{\mathrm{min}}$, the oscillation amplitude $A$ and the bandwidth $B$ were set to $100 \mathrm{~pm}$ and $5 \mathrm{~kHz}$, respectively. 
The results are shown in Figures 3(a) and 3(b).
It can be seen that both $n_{\mathrm{ds}}$ and $F'_{\mathrm{min}}$ can be reduced by shortening $L$. 
However, when $C_{1}$ is large, the reduction rate of $F'_{\mathrm{min}}$ remains small even if $L$ is shortened. 
In other words, decreasing both $C_{1}$ and $L$ is crucial for the effective reduction of $n_{\mathrm{ds}}$ and $F'_{\mathrm{min}}$.
This can be explained as follows. 
As shown in Fig. 1(c), $I_{\mathrm{n}}$ is dominated by $i_{\mathrm{n}}$ in the low-frequency region and by $e_{\mathrm{n}}$ in the high-frequency region. 
In the $e_{\mathrm{n}}$-dominant region, $I_{\mathrm{n}}$ increases as the frequency rises. 
The crossover frequency at which the dominant noise source shifts from $i_{\mathrm{n}}$ to $e_{\mathrm{n}}$ depends on $C_{1}$, and decreasing $C_{1}$ can shift this frequency to a higher range. 
That is, reducing $C_{1}$ is significant for two reasons: it lowers the absolute value of $I_{\mathrm{n}}$ and suppresses the increase of $I_{\mathrm{n}}$ relative to frequency increments.

\begin{figure}[htb]
\includegraphics{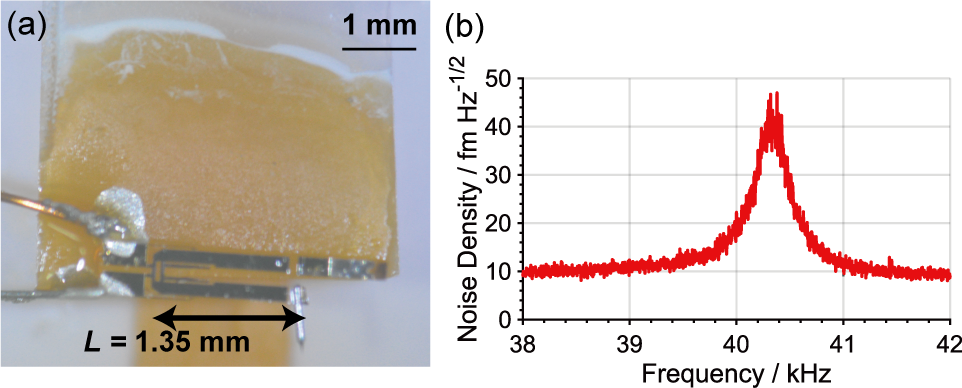}
\caption{\label{Fig_4} (a) Photograph of a newly fabricated qPlus sensor with a shortened prong and a porous-glass substrate. (b) Thermal Brownian Spectrum of the fabricated qPlus sensor, $f = 40330 \mathrm{~Hz}$, $k = 5543 \mathrm{~N~m^{-1}}$, $Q =141$, $S_\mathrm{v} = 127.3 \mathrm{~\mu V~m^{-1}}$, $n_{\mathrm{ds}} = 9.3 \mathrm{~fm~Hz^{-1/2}}$.}
\end{figure}

Based on the discussions above, we fabricated a qPlus sensor using a porous glass substrate and a low-$\epsilon$ epoxy adhesive (323LP), with the QTF prongs shortened. 
A photograph of the sensor is shown in Fig. 4(a). 
Almost the entire electrode-free portion of the QTF was removed, resulting in $L = 1.35 \mathrm{~mm}$ relative to the electrode length $L_e = 1.16 \mathrm{~mm}$.
The yellow coloration of the substrate is due to the penetration of the epoxy into the pores. 
However, even assuming that the pores are completely filled with epoxy, it can be concluded that a lower dielectric constant is maintained compared to conventional alumina substrates, considering the dielectric constant of glass ($\epsilon \simeq 4$) and the porosity of the substrate.
Figure 4(b) shows the thermal oscillation spectrum of the fabricated qPlus sensor. 
The $n_{\mathrm{ds}}$ was measured to be $9.3 \mathrm{~fm~Hz^{-1/2}}$, which represents a successful reduction to one-third of the conventional value. 
This experimentally demonstrates that the theory and methods proposed in this paper are effective for reducing $n_{\mathrm{ds}}$. 
However, the $Q$ factor was found to be 141, which is significantly lower than typical values for qPlus sensors. 
We speculate that this is due to the introduction of defects into the quartz during the shortening of the QTF prong. 
Such defects cause internal friction and increase energy dissipation, thereby decreasing the $Q$ factor. 
Since a lower $Q$ factor leads to increased noise in the low-frequency region, improvements in the cutting method are desired.
However, since $n_{\mathrm{ds}}$ is dominant in the frequency range of high-speed AFM ($> 1 \mathrm{~kHz}$), it is suggested that the proposed method is highly effective for improving overall noise performance.
When $F'_{\mathrm{min}}$ for this sensor is calculated according to Eq. (1) under the conditions of $B = 5 \mathrm{~kHz}$ and $A = 100 \mathrm{~pm}$, the result is $F'_{\mathrm{min}} \sim 5 \mathrm{~N/m}$. 
In contrast, the $F'_{\mathrm{min}}$ for a conventional qPlus sensor is estimated to be $\sim 10 \mathrm{~N/m}$ as shown in Figure 1(a).
That is, we have not only reduced $n_{\mathrm{ds}}$ to one-third of its original value but also succeeded in reducing $F'_{\mathrm{min}}$ by half.

\begin{figure*}[htb]
\includegraphics{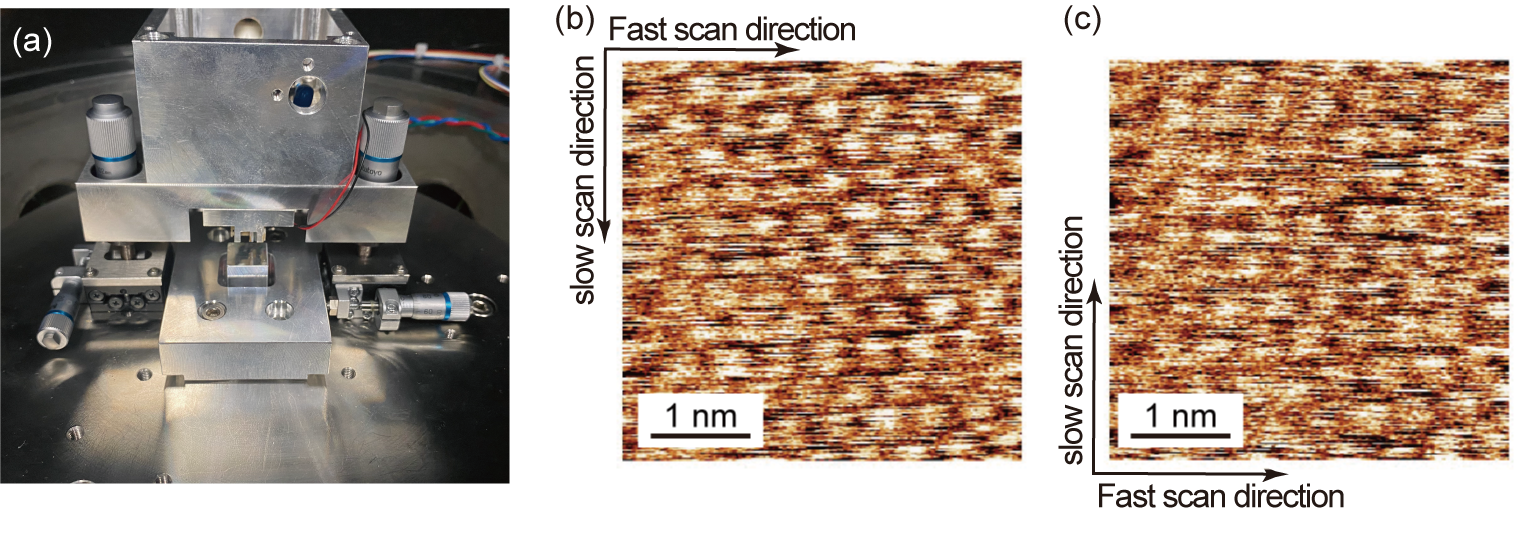}
\caption{\label{Fig_5} (a) Photograph of the developed sample-scanning AFM system. (b), (c) Sequential AFM topographic images of an $\mathrm{AuGa}_{2}(111)$ surface in molten gallium ($4\mathrm{~nm} \times 4\mathrm{~nm}$, $256 \times 256$ pixels). The scanning time was $6.6\mathrm{~s}$ per frame (39 lines $\mathrm{s^{-1}}$). $f = 45928 \mathrm{~Hz}$, $k = 6705 \mathrm{~N~m^{-1}}$, $Q =158$, $S_\mathrm{v} = 114.8 \mathrm{~\mu V~m^{-1}}$, $n_{\mathrm{ds}} = 12.5 \mathrm{~fm~Hz^{-1/2}}$.}
\end{figure*}

Finally, we performed atomic-resolution AFM analysis in a liquid environment using the fabricated qPlus sensor. 
We developed a sample-scanning AFM system designed to minimize $C_{1}$ by shortening the wiring length between the qPlus sensor and the TIA (Figure 5(a)). 
A Quadpod scanner with a high resonance frequency was employed to facilitate high-speed scanning\cite{Nishiwaki26}. 
The frequency demodulation bandwidth ($B$) was set to 5 $\mathrm{~kHz}$ using a Zurich Instruments MFLI lock-in amplifier and a laboratory-developed hybrid-loop PLL\cite{Nishiwaki26}. 
The sample was prepared by depositing $200 \mathrm{~nm}$ of gold onto a mica substrate, followed by the deposition of a molten gallium droplet, which was then left at room temperature for over 12 hours. 
The electronic control of the AFM was performed by a Nanonis BP5e SPM control system (Nanonis-SPECS Zurich GmbH, Zurich, Switzerland).
AFM analysis of the molten gallium/Au-Ga alloy interface was conducted by immersing the tungsten tip of the qPlus sensor into the molten gallium.
Figures 5(b) and 5(c) show sequential topographic images obtained over a $4 \mathrm{~nm} \times 4 \mathrm{~nm}$ area. 
Each image consists of $256 \times 256$ pixels, with a scanning time of $6.6 \mathrm{~s}$ (39 lines s$^{-1}$). 
In both images, the fast scan direction is from left to right, while the slow scan direction is from top to bottom for (b) and bottom to top for (c). 
It should be noted that no post-processing for noise reduction was applied to these images. 
In both cases, bright spots with six-fold symmetry and a period of approximately $0.5 \mathrm{~nm}$ were clearly resolved. 
These results are in good agreement with the crystal structure of $\mathrm{AuGa}_{2}(111)$\cite{Godwal13} and consistent with our previous study\cite{Ichii21}. 
Thus, it was experimentally demonstrated that atomic-resolution analysis is feasible even at such high scanning speeds. 
Furthermore, despite the inversion of the slow scan direction, the atomic arrangement in (b) and (c) remains almost identical. 
This indicates that the influence of thermal drift was sufficiently suppressed by the high-speed scanning, highlighting the practical significance of our high-speed AFM approach.

In summary, we developed a low-noise qPlus sensor optimized for high-speed AFM in liquid environments. 
By analyzing the current noise density $I_{\mathrm{n}}$ and its frequency dependence, we demonstrated that reducing the input capacitance $C_{1}$ and shortening the QTF prongs are essential for minimizing the displacement sensor noise density $n_{\mathrm{ds}}$. 
The developed sensor, featuring a porous glass substrate and shortened prongs, achieved an $n_{\mathrm{ds}}$ of $9.3 \mathrm{~fm~Hz^{-1/2}}$, approximately one-third that of conventional sensors. 
Furthermore, the minimum detectable force gradient $F_{\mathrm{min}}^{\prime}$ at $B = 5$ kHz was reduced by half. 
Finally, we demonstrated the practical performance of our sensor through atomic-resolution imaging of a molten gallium/Au-Ga alloy interface at a high scanning speed of 39 lines $\mathrm{s^{-1}}$. 
The demonstrated performance significantly extends the capabilities of qPlus-based AFM, facilitating high-speed, atomic-scale research in various liquids such as molten metals and polymer melts where conventional methods face limitations.

This work was supported by JSPS KAKENHI for  for Scientific Research B (JP23K26543) and JST PRESTO (JPMJPR25J2).

\section* {AUTHOR DECLARATIONS}
\subsection* {Conflict of Interest}
The authors have no conflicts to disclose.

\subsection* {Author contribution}
T. I. coordinated the project and drafted the original paper. 
S. T. and Y. N performed the AFM experiment.
T. U. contributed to the interpretation of the results.
All authors discussed the results and contributed to the preparation of the paper.



\end{document}